\documentclass[%
 reprint,
superscriptaddress,
nofootinbib,
 amsmath,amssymb,
]{revtex4-2} 

\usepackage{graphicx}
\usepackage{xcolor} 
\usepackage{dcolumn}
\usepackage[english]{babel}
\usepackage[toc,page]{appendix}
\usepackage{bm}
\usepackage{hyperref}

\begin{document}

\preprint{APS/123-QED}

\title{Introduction to Higher Order Classical Dynamics: Pais-Uhlenbeck Model and Coupled Oscillators}

\author{C\'{a}ssius Anderson Miquele de Melo }\thanks{cassius@unifal-mg.edu.br} \homepage{https://orcid.org/0000-0001-5096-1297}
\address{Instituto de Ci\^{e}ncia e Tecnologia, Universidade Federal de Alfenas, BR 267 - Rodovia José Aur\'{e}lio Vilela, nº 11.999, Km 533 37715-400 Cidade Universit\'{a}ria, Poços de Caldas, Minas Gerais, Brasil.
}
\address{INFN, Laboratori Nazionali del Sud (LNS), Via S. Sofia 62, 95123 Catania, Italy}
\address{Universit\`{a} di Catania, Dipartimento di Fisica e Astronomia “Ettore Majorana” (INFN-CT), Via Santa Sofia 64, 95123 Catania, Italy}

\author{Ivan Francisco de Souza}\thanks{ivanfrancisco2098@gmail.com}\homepage{https://orcid.org/0000-0002-1679-0057}
\address{Instituto de Ci\^{e}ncia e Tecnologia, Universidade Federal de Alfenas, BR 267 - Rodovia José Aur\'{e}lio Vilela, nº 11.999, Km 533 37715-400 Cidade Universit\'{a}ria, Poços de Caldas, Minas Gerais, Brasil.
}

\date{\today}

\begin{abstract}

{\bf Abstract:} Most of the laws of Nature involve derivatives up to the second order. Ostrogradski was the first to seek a formulation of the equations of higher-order derivatives. He extended Hamilton's equations by considering Lagrangians that depend on higher-order derivatives of generalized coordinates. The Hamilton-Ostrogradski formulation served as the basis for later studies with higher-order derivatives. However, the Hamilton-Ostrogradski formalism is rarely discussed in textbooks or the pedagogical literature. This motivated us to show how the Hamilton-Ostrogradski formalism can be applied it to the Pais-Uhlenbeck oscillator. We hope that the approach presented in this work can serve as a basis for discussion in advanced classical mechanics courses.
\end{abstract}

\keywords{Suggested keywords}
\maketitle


\section{\label{sec:level1}Introduction\protect\\}
 Many physics equations involve only first- and second-order derivatives, such as Newton's laws, Maxwell's equations and Schrödinger’s equation. However, nothing prevents physical systems in nature from being dependent on higher-order derivatives.

There are theoretical models that involve higher-order derivatives, such as Generalized Electrodynamics \cite{podolsky1942generalized, cuzinatto2011can}, which is a particular case of higher-order gauge theories \cite{cuzinatto2007second}. One of the main motivations to study higher derivative theories is that they represent high-energy corrections to our known low-energy effective theories of nature. This is connected to the Wilsonian/Renormalisation group interpretation of quantum field theories \cite{shapiro2008effective, borges2019higher}. Another common case of theories with higher-order derivatives is gravity \cite{stelle1978classical, lu2015spherically, cuzinatto2008gauge, Sotiriou2008fRTOA, cuzinatto2011cosmic, cuzinatto2015observational}, whose higher-order derivatives can be seen as evidence of coupling with new fields beyond the standard model \cite{cuzinatto2016scalar, cuzinatto2019f}. Furthermore, there is practical interest in higher order derivatives for quantum gravity and field theory, as such derivatives naturally arise as effective corrections and candidates for fundamental theories, such as string theory \cite{green1987, zwiebach2009, Polchinski:1998rq, Mendes2017, eliezer1989problem}.

There are also concrete physical examples of higher-order derivative forces. A well-known case is the Abraham–Lorentz force \cite{Abraham1905,Lorentz1909, dirac1938classical} in classical electrodynamics \cite{Jackson1999,spohn2004dynamics,Rohrlich2007}. This force describes the reaction of a charged particle that emits radiation when accelerated. In this situation, the force depends on the derivative of the acceleration (often called the “jerk”), leading to equations of motion that are third\textcolor{violet}{-}order in time. Such dynamics give rise to peculiar effects such as preacceleration, where the particle starts to move before the external force is applied, and runaway solutions. These features illustrate that generalizations of Newton’s second law to higher orders are indeed possible, but they come with important conceptual and practical challenges.

The first and perhaps most influential example of a higher-order system in classical mechanics is the Pais–Uhlenbeck oscillator, originally introduced in 1950 as a model for field theories with non-localized action \cite{Pais1950}. Since then, it has become a paradigmatic system for exploring the conceptual challenges associated with higher-order dynamics. In particular, it embodies the so-called Ostrogradsky instability, the hallmark of unbounded Hamiltonians in higher-derivative theories \cite{woodard2015theorem}. Furthermore, the Pais-Uhlenbeck oscillator has several modern applications, such as the description of ion traps \cite{Guha2020CurlFA}, circularly polarized gravitational waves \cite{Elbistan2022CircularlyPP}, and dark energy \cite{comelli2022classical}. These features make the Pais–Uhlenbeck oscillator not only historically relevant but also a valuable pedagogical tool to motivate the study of higher-order formalisms.

To describe physical systems with higher-order derivatives, it is necessary to have equations involving derivatives of any order. The first person to formulate equations of motion for higher orders was Mikhail Vassilievich Ostrogradsky \cite{Ostrogradski1850}. In 1850, Ostrogradsky derived the Hamiltonian equations when considering Lagrangians that depend on higher-order time derivatives of generalized coordinates \cite{Ostrogradski1850}.

Equations of motion involving higher-order derivatives are a subject often overlooked in various textbooks and advanced mechanics courses \cite{Goldstein2002, arnol2013mathematical}. In recent decades, this topic has been extensively studied, as some researchers believe that new discoveries in this area could expand our understanding of nature \cite{caro2021metodo}. However, it remains relatively unknown among many physics students. With this in mind, this work aims to provide a brief introduction to higher-order equations of motion, hoping to serve as motivation and a foundation for students who wish to delve deeper and consult other references covering this subject.

In Section II, we present the Lagrange equations for a Lagrangian that depends on the $n$-th time derivative of the generalized coordinates. In Section III, we discuss the transition from the Lagrangian to the Hamiltonian formalism in higher-order theories. Our aim is to apply this formalism to the Pais-Uhlenbeck oscillator, but since it is not common to deal with energies depending on second-order derivatives, understanding the Pais-Uhlenbeck oscillator may initially be challenging. For this reason, in Section IV we introduce the model of two coupled oscillators, which is easier to visualize and which, when reformulated to yield uncoupled fourth-order equations of motion, can be mapped onto the Pais-Uhlenbeck model. Finally, Section V is devoted to the Pais-Uhlenbeck oscillator, whose Lagrangian involves derivatives up to the second order and leads to a fourth-order equation of motion.  In Section VI, we present our final considerations.

\section{Higher-Order Lagrange Equations}

Newton's second law is based on empirical observations of nature rather than a general mathematical principle. Therefore, it is not straightforward to extend it systematically to higher-order derivatives of position. In contrast, Lagrange's and Hamilton's equations can be derived from the principle of least action, providing a consistent mathematical framework for constructing higher-order equations.

In the literature and in advanced mechanics courses, usually only the first-order Lagrangian formulation is presented, $L=L(q,\dot{q},t)$, where $q\equiv q(t)$ is the generalized coordinate, $\dot{q}\equiv \mathrm{d}q/\mathrm{d}t$ is the generalized velocity (first-order derivative) and $t$ is time.

The Lagrange equation for this type of Lagrangian is

\begin{equation}
\label{E2}
\frac{\partial L}{\partial q}-\frac{\mathrm{d} }{\mathrm{d} t}\left(\frac{\partial L}{\partial \dot{q}}\right)=0.
\end{equation}

Observe that this is an equation that involves derivatives up to the second order, which can be seen by expanding the second term of Eq.\eqref{E2}. This equation is useful for the vast majority of physical systems, but for systems that might depend on derivatives of order greater than two, it is necessary to generalize the Lagrange equations.

Going further, considering a Lagrangian that can also depend on $\ddot{q}$, the generalized acceleration, by imposing certain conditions and performing some mathematical procedures (see Appendix A), one obtains the following Lagrange equation:

\begin{equation}
\label{E3}
\frac{\partial L}{\partial q}-\frac{\mathrm{d} }{\mathrm{d} t}\left(\frac{\partial L}{\partial \dot{q}}\right)+\frac{\mathrm{d}^{2} }{\mathrm{d} t^{2}}\left(\frac{\partial L}{\partial \ddot{q}}\right)=0.
\end{equation}

Note that now it is an equation that may involve derivatives up to the fourth order. It is possible to generalize the Lagrange equations that depend on the $n$-th order derivative. The higher-order Lagrange equation is
\begin{equation}
\label{E4}
\sum_{k=0}^{n}(-1)^{k}\frac{\mathrm{d}^{k} }{\mathrm{d} t^{k}}\left ( \frac{\partial L}{\partial q^{(k)}} \right )=0.
\end{equation}

We write the index indicating the order of the derivative in parentheses “( )” to avoid confusion with exponents. For example, $(-1)^{k}$ means $-1$ raised to the $k$-th power, while $q^{(k)}$ refers to the $k$-th order derivative of the generalized coordinate. For $n=1$, the higher-order equation reduces to Eq.\eqref{E2}.

If the system has $N$ degrees of freedom, there will be $N$ generalized coordinates and $N$ higher-order Lagrange equations:
\begin{equation}
\label{E5}
\sum_{k=0}^{n}(-1)^{k}\frac{\mathrm{d}^{k} }{\mathrm{d} t^{k}}\left ( \frac{\partial L}{\partial q_{i}^{(k)}} \right )=0 \quad (i=1,...,N).
\end{equation}

Eq.\eqref{E5} is an equation that may involve derivatives up to order $2n$. 
We can define a matrix, called the \textit{Hessian}, which plays a central role in the analysis of higher-order systems. Its elements are given by  

\begin{equation}
W_{ij}=\frac{\partial^2 L}{\partial q_{i}^{(n)}\partial q_{j}^{(n)}} \quad (i,j=1,\dots,N).
\end{equation}

If this matrix is non-singular, that is, if its determinant is nonzero, it becomes possible to isolate the highest-order terms $q_{i}^{(2n)}$ from the Lagrange equation, yielding  

\begin{equation}
W_{ij}q_{j}^{(2n)}=F_{i}(q_{i},...,q_{i}^{(2n-1)},t) \quad (i,j=1,...,N),
\end{equation}
where $F_{i}(q_{i},...,q_{i}^{(2n-1)},t)$ collects all the terms involving derivatives of lower order.  

In this work, we deal only with regular systems. In the case of systems where the Hessian matrix is singular, one of the established methods to construct the Hamiltonian formulation must be applied \cite{caro2021metodo, bertin2008formalismo, 10.1119/5.0107540}.

\section{Higher-Order Hamilton Equations}

In this section, we will review Hamilton's equations as they are usually presented, considering the Lagrangian of Eq.~\eqref{E2}. We will then consider a Lagrangian with second-order derivatives and, present the general formulation of Hamilton's equations.

As seen in the previous section, Eq.~\eqref{E2} contains derivatives only up to the second order, which is sufficient for many physical systems. The advantage of moving from the Lagrangian formalism to the Hamiltonian formalism is that Hamilton's equations are first-order equations. In return, the number of variables and equations is doubled, since generalized momenta are introduced, defined by
\begin{equation}
\label{E8}
p_{i}=\frac{\partial L}{\partial \dot{q}_{i}} \quad (i=1,...,N).
\end{equation}

To construct the Hamiltonian function, the Legendre transformation is used:
\begin{equation}
\label{E9}
H(q_{i},p_{i},t)=\sum_{i=1}^{N}p_{i}\dot{q}_{i}-L(q_{i},\dot{q}_{i},t).
\end{equation}

Again, the condition that the Hessian matrix is regular must be satisfied so that it is possible to invert the $N$ equations of the type Eq.~\eqref{E8} and express the $\dot{q}_{s}$ as functions of $q_{s}$, $p_{s}$, and $t$. Thus, the Hamiltonian formulation has a set of $2N$ equations that can be solved to determine the $2N$ independent variables. The usual Hamilton equations are:
\begin{equation}
\label{E10}
\begin{aligned}
\dot{q}_{i}=\frac{\partial H}{\partial p_{i}},\\
\dot{p}_{i}=-\frac{\partial H}{\partial q_{i}}.
\end{aligned}
\end{equation}

When considering higher-order Lagrangians, the Legendre transformation as presented in Eq.~\eqref{E9} does not generate the corresponding Hamilton equations for higher orders~\cite{Rashid1996}. The transition between formalisms when higher-order derivatives are considered is somewhat more subtle, as new coordinates and momenta must be defined. We will see that such definitions will imply constraints, which are unavoidable when considering Lagrangians depending on derivatives of order greater than one, and because of that, not all variables will be mutually independent. Therefore, a more general Legendre transformation is necessary for higher-order equations. The first to obtain the Hamiltonian formulation for higher-order Lagrangians was Ostrogradsky~\cite{Ostrogradski1850}. This formulation became known as the Hamilton-Ostrogradsky formalism.

We will take a simpler approach to the Hamilton-Ostrogradsky formalism, considering that the system has only one degree of freedom and that its Lagrangian does not explicitly depend on time. Considering the case of a Lagrangian of the type $L(q,q^{(1)},q^{(2)})$, we saw that the Lagrange equation takes the form of Eq.~\eqref{E3}. Firstly, the canonical variables introduced by Ostrogradsky, for a Lagrangian that depends up to the second time derivative of \(q(t)\), are defined as
\begin{equation}
\label{E11}
\begin{split}
Q_{1}&=q,\\
Q_{2}&=q^{(1)},\\ 
P_{1}&=\frac{\partial L}{\partial \dot{Q_1}}-\frac{\mathrm{d} }{\mathrm{d} t}\left ( \frac{\partial L}{\partial \ddot{Q_1}} \right ),\\
P_{2}&=\frac{\partial L}{\partial \dot{Q}_2},
\end{split}
\end{equation}  
where \(Q_{1}\) and \(Q_{2}\) are the new generalized coordinates, and \(P_{1}\) and \(P_{2}\) are the corresponding conjugate momenta.

These definitions are not arbitrary: mathematically, the purpose of Ostrogradsky’s formalism is to recast the dynamics of higher-order systems into a set of first-order differential equations. For a Lagrangian of the form \(L=L(q,\dot{q},\ddot{q})\), the corresponding Euler-Lagrange equation is of fourth order in \(q(t)\). By introducing the variables \((Q_{1},Q_{2},P_{1},P_{2})\), the problem is reformulated in an extended phase space, where the temporal evolution can be described by the standard Hamiltonian formalism, although in a higher dimension.

From a physical point of view, \(Q_{1}=q\) represents the position, while it \(Q_{2}=\dot{q}\) corresponds to the velocity promoted to an independent coordinate, necessary to describe the dynamics of a fourth-order system. The momenta also acquire an enriched interpretation: \(P_{2}=\partial L/\partial \ddot{q}\) plays the role of the momentum conjugate to the velocity, capturing the sensitivity of the Lagrangian to the acceleration, while \(P_{1}\) includes not only the usual contribution \(\partial L/\partial \dot{q}\), but also an additional term coming from the time variation of \(\partial L/\partial \ddot{q}\). This demonstrates how the canonical momentum in higher-order theories encodes the system's reaction to variations in generalized velocities rather than just being the rate at which the Lagrangian varies in relation to velocity

 Notice that, due to the definition of momentum, constraints arise, such as:
\begin{equation}
\label{E12}
P_{1}+\dot{P}_{2}-\frac{\partial L}{\partial q^{(1)}}=0.
\end{equation}

To obtain the canonical equations, it is necessary that the Hessian matrix be regular so that the highest-order derivative can be written in terms of the new variables,\footnote{In this case, $q^{(2)}$ as a function of $Q_{1}$, $Q_{2}$, and $P_{2}$.} thus allowing the Hamiltonian to be written for this case (cf. Appendix B). Otherwise, it would be necessary to use the Dirac-Bergmann method \cite{10.1119/5.0107540} to handle the constraints that arise from these definitions. We will not address the Dirac-Bergmann method here; instead, we will consider cases where the Hessian matrix is regular. After some mathematical procedures (cf. Appendix B), the Legendre transformation obtained is:
\begin{equation}
\label{E13}
\begin{aligned}
H(Q_{1},Q_{2},P_{1},P_{2})=P_{1}\dot{Q}_{1}+P_{2}\dot{Q_2}(Q_{1},Q_{2},P_{2})\\
-L(Q_{1},Q_{2}, P_{1},P_{2},q^{(2)}(Q_{1},Q_{2},P_{2})).
\end{aligned}\,
\end{equation}

Thus, the Hamilton equations \textcolor{blue}{are}:
\begin{equation}
\begin{aligned}
\label{E14}
\dot{Q}_{1}=\frac{\partial H}{\partial P_{1}},\\
\dot{Q}_{2}=\frac{\partial H}{\partial P_{2}},\\
\dot{P}_{1}=-\frac{\partial H}{\partial Q_{1}},\\
\dot{P}_{2}=-\frac{\partial H}{\partial Q_{2}}.
\end{aligned}\,
\end{equation}

Even considering only one degree of freedom, when transitioning to the Hamiltonian formulation, four variables are defined, requiring four Hamilton equations. Appendix B shows how these results are obtained. It is important to emphasize that these equations can only be derived if a non-degenerate Lagrangian is considered, that is, one whose Hessian matrix is invertible.

The most general form used by Ostrogradsky to define the canonical variables is:
\begin{equation}
\label{E15}
Q_{k}=q^{(k-1)} \quad (k=1,...,n),
\end{equation}
\begin{equation}
\label{E16}
P_{k}=\sum_{l=k}^{n}(-1)^{l-k}\frac{\mathrm{d} ^{l-k}}{\mathrm{d} t^{l-k}}\left ( \frac{\partial L}{\partial q^{(l)}} \right ) \quad (k=1,...,n),
\end{equation}
where $Q_{k}$ is the $k$-th coordinate and $P_{k}$ is the $k$-th momentum. A total of $2n$ variables are defined, although not all of them are independent because of the constraints arising from these definitions. The general Legendre transformation has the following form:
\begin{equation}
\label{E17}
H=\sum_{k=1}^{n-1}P_{k}Q_{k+1}+P_{n}q^{(n)}-L=\sum_{k=1}^{n}P_{k}\dot{Q}_{k}-L,
\end{equation}
and the $2n$ Hamilton equations are:
\begin{equation}
    \label{E18}
    \begin{aligned}
        \dot{Q}_{k}=\frac{\partial H}{\partial P_{k}},\\
\dot{P}_{k}=-\frac{\partial H}{\partial Q_{k}}.
    \end{aligned}\, \quad (k=1,...,n)
\end{equation}

For a system with $N$ degrees of freedom, each of them of order $n$, there will be a total of $N\times 2n$ equations for $N\times 2n$ variables:
\begin{equation}
\label{E19}
    \begin{aligned}
        \dot{Q}_{i,k}=\frac{\partial H}{\partial P_{i,k}},\\
        \dot{P}_{i,k}=-\frac{\partial H}{\partial Q_{i,k}},
    \end{aligned}\,
\end{equation}
where $i=1,...,N$ and $k= 1,...,n$.

\section{Coupled Oscillators}

An interesting model that can be used as a first approach to higher-order equations is that of two coupled oscillators. We will see that it is not necessary to resort to higher-order Lagrange or Hamilton–Ostrogradsky equations to obtain the equations of motion for these oscillators, since the Lagrangian of this system depends only on first-order derivatives. Nevertheless, it can also be shown that the solutions of the second-order equations of motion for the coupled oscillators form a subset of the solutions of the fourth-order equation of motion of the Pais–Uhlenbeck oscillator, for which higher-order equations are required \cite{kleefeld2023equivalence}.

To begin the discussion, consider two oscillators of masses $m_{1}$ and $m_{2}$ attached to walls by springs of constants $k_{1}$ and $k_{2}$, respectively; between the two masses there is a third spring of constant $k$ coupling them, as shown in Figure \ref{fig:Fig.1}:

\begin{figure}[h]
    \centering
    \includegraphics[scale=0.17]{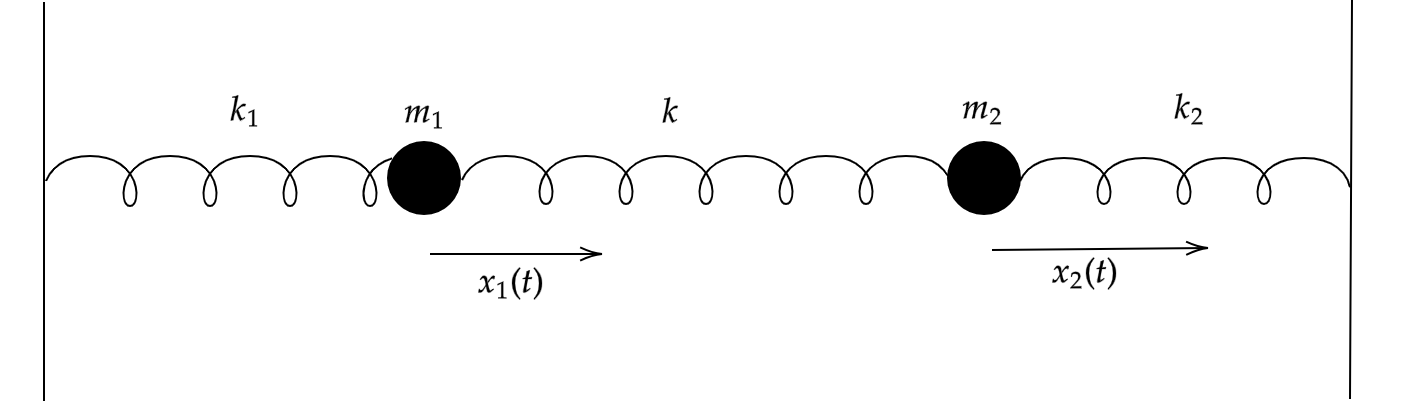}
    \caption{Schematic diagram of two coupled oscillators}
    \label{fig:Fig.1}
\end{figure}

Let $x_{1}(t)$ and $x_{2}(t)$ denote the displacements of masses $m_{1}$ and $m_{2}$ from their respective equilibrium positions.
 The Lagrangian of the coupled oscillators is
\begin{equation}
    \label{E20}
    L=\frac{m_{1}\dot{x}_{1}^{2}}{2}+\frac{m_{2}\dot{x}_{2}^{2}}{2}-\frac{k_{1}x_{1}^{2}}{2}-\frac{k_{2}x_{2}^{2}}{2}-\frac{k(x_{2}-x_{1})^{2}}{2}.
\end{equation}

This is a system with two degrees of freedom. In this case, we can use Eq.\eqref{E2} to obtain the equations of motion:
\begin{equation}
    \label{E21}
    \left\{
    \begin{aligned}
        m_{1} \ddot{x}_{1} + k_{1} x_{1} - k(x_{2} - x_{1}) &= 0, \\
        m_{2} \ddot{x}_{2} + k_{2} x_{2} - k(x_{1} - x_{2}) &= 0.
    \end{aligned}
    \right.
\end{equation}

Note that in these two equations of motion the oscillators’ motions are not independent due to the coupling constant $k$. In the first equation (oscillator 1) the coordinate of oscillator 2 appears, and vice versa in the second equation. To find the solutions of these two equations, one method is the order-raising technique: differentiate one equation twice and substitute the other into it, yielding a single fourth-order equation involving only one oscillator’s coordinate. Applying this method (see Ref.\cite{Mendes2017}), the pair in Eq.\eqref{E21} reduces to:
\begin{equation}
    \label{E22}
    \begin{aligned}
    m_{1}\ddddot{x_1}+\left[\frac{m_2(k_{1}+k)+m_1(k_{2}+k)}{m_2}\right]\ddot{x}_{1}+\\
    +\left[\frac{k_1k_2+k(k_1+k_2)}{m_2}\right]x_{1}=0,
    \end{aligned}
\end{equation}

The fourth-order equation for oscillator 2 is analogous, obtained by swapping indices 1 and 2 in Eq.\eqref{E22}.

The general solution of Eq.\eqref{E22} is
\begin{equation}
    \label{E23}
    \begin{aligned}
    x_{1}(t)=A_{1}\cos(\omega_{1}t)+B_{1}\sin(\omega_{1}t)+\\
    +A_{2}\cos(\omega_{2}t)+B_{2}\sin(\omega_{2}t),
    \end{aligned}
\end{equation}
where $\omega_{1}$ and $\omega_{2}$ are
\begin{equation}
    \label{E24}
    \omega_1=\sqrt{\frac{\alpha+\beta}{2}} \quad \mathrm{e}\quad \omega_2=\sqrt{\frac{\alpha-\beta}{2}}
\end{equation}
in which, $\alpha$ and $\beta$ are defined as follows:
\begin{equation*}
    \alpha\equiv \frac{k_{2}+k}{m_{2}}
  + \frac{k_{1}+k}{m_{1}},
\end{equation*}
\begin{equation*}
    \beta \equiv\sqrt{
      \left(
        \frac{k_{2}+k}{m_{2}}
      - \frac{k_{1}+k}{m_{1}}
      \right)^{2}
      - \frac{4k^{2}}{m_{1}m_{2}}
    }.
\end{equation*}

The constants $A_{1}$, $A_{2}$, $B_{1}$, and $B_{2}$ depend on the initial conditions:
\begin{equation}
    \label{E25}
    \begin{aligned}
A_{1}=\frac{\omega_{2}^{2}x_{1}(0)+\ddot{x}_{1}(0)}{\omega_{2}^{2}-\omega_{1}^{2}},\\
A_{2}=-\frac{\omega_{1}^{2}x_{1}(0)+\ddot{x}_{1}(0)}{\omega_{2}^{2}-\omega_{1}^{2}},\\
B_{1}=\frac{\omega_{2}^{2}\dot{x}_{1}(0)+\dddot{x}_{1}(0)}{\omega_{1}(\omega_{2}^{2}-\omega_{1}^{2})},\\
B_{2}=-\frac{\omega_{1}^{2}\dot{x}_{1}(0)+\dddot{x}_{1}(0)}{\omega_{2}(\omega_{2}^{2}-\omega_{1}^{2})}.
\end{aligned}
\end{equation}

Thus, by using this method, the two second-order equations reduce to one fourth-order equation. Once $x_{1}(t)$ is determined, one isolates $x_{2}$ from the first line of Eq.\eqref{E21} and substitutes $x_{1}(t)$ and $\ddot{x}_{1}(t)$ to find the solution for oscillator 2.

In the next section, we will present the Lagrangian of the Pais–Uhlenbeck oscillator, which yields a fourth-order equation of motion. We will also show that the solutions for the coupled oscillators satisfy the Pais–Uhlenbeck oscillator’s equation of motion.

\section{Pais-Uhlenbeck Oscillator}

The Pais-Uhlenbeck oscillator is a model that yields extensive analysis, with one particular case being that of coupled oscillators. Therefore, in this section, we will focus on discussing how the two-coupled-oscillators model seen in the previous section is equivalent to the Pais-Uhlenbeck oscillator.

The Lagrangian of the Pais-Uhlenbeck oscillator is
\begin{equation}
    \label{E26}
    L_{\mathrm{PU}}=\frac{1}{2}\left[\ddot{x}^{2}+(\Omega_{1}^{2}+\Omega_{2}^{2})\dot{x}^{2}+\Omega_{1}^{2}\Omega_{2}^{2}x^{2}\right],
\end{equation}
where $\Omega_{1}$ and $\Omega_{2}$ are constants.

This is a Lagrangian that depends on a second-order derivative, which requires the use of a higher-order Lagrange equation to obtain the equation of motion. Applying Eq.\eqref{E3} to this Lagrangian, we obtain:
\begin{equation}
    \label{E27}
    \ddddot{x}+(\Omega_{1}^{2}+\Omega_{2}^{2})\ddot{x}+\Omega_{1}^{2}\Omega_{2}^{2}x=0.
\end{equation}

We have a fourth-order equation of motion. If we divide Eq.\eqref{E22} by $m_{1}$, we can notice the similarity between Eq.\eqref{E22} and Eq.\eqref{E27}. For them to be equivalent, the following relations must be satisfied:
\begin{equation}
    \label{E28}
    \begin{aligned}
        \Omega_{1}^{2}+\Omega_{2}^{2}=\frac{k_{1}+k}{m_{1}}+\frac{k_{2}+k}{m_{2}},\\
        \Omega_{1}^{2}\Omega_{2}^{2}=\frac{k_{1}k+k_{2}k+k_{1}k_{2}}{m_{1}m_{2}}.
    \end{aligned}
\end{equation}

Solving these two equations to determine $\Omega_{1}$ and $\Omega_{2}$, we have

\begin{equation}
    \label{E29}
    \Omega_1=\sqrt{\frac{\alpha+\beta}{2}} \quad \mathrm{e}\quad \Omega_2=\sqrt{\frac{\alpha-\beta}{2}}
\end{equation}

The right-hand sides of Eq.\eqref{E29} and Eq.\eqref{E24} are exactly the same, implying that $\Omega_{1}=\omega_{1}$ and $\Omega_{2}=\omega_{2}$. Since $x_{1}(t)$ and $x_{2}(t)$ are solutions of the coupled oscillators, we have an equivalence between the Pais-Uhlenbeck oscillator and the coupled oscillators:
\begin{equation}
    \label{E30}
    L=\frac{1}{2}\left[\ddot{x}_{i}^{2}+(\omega_{1}^{2}+\omega_{2}^{2})\dot{x}_{i}^{2}+\omega_{1}^{2}\omega_{2}^{2}x_{i}^{2}\right],
\end{equation}
where $i=\{1,2\}$ so that each one of the coupled oscillators can be taken as an unique Pais-Uhlenbeck oscillator.

Despite the formal equivalence of the Lagrangians, it is crucial to draw attention to some nuances related to the order elevation method. First, the fourth-order equations require a larger number of initial conditions than the original second-order system, which gives rise to additional nonphysical (spurious) solutions that must be properly identified and discarded through an analysis of the boundary conditions. Furthermore, in higher-order formulations, the variational principle naturally imposes boundary conditions, and translating these into initial conditions is not always straightforward, requiring careful consideration \cite{mukhanov1995symmetries}. A more detailed analysis of these aspects lies beyond the scope of the present work but could be pursued in future studies. 

For a Hamiltonian approach to the Pais-Uhlenbeck oscillator, we must first write the new coordinates and momenta from the Ostrogradsky definition. Through Eq.\eqref{E11}, we have:
\begin{equation}
\label{E31}
\begin{split}
Q_{1}&=x_{1},\\
Q_{2}&=\dot{x}_{1},\\ P_{1}&=\frac{\partial L}{\partial q^{(1)}}-\frac{\mathrm{d} }{\mathrm{d} t}\left ( \frac{\partial L}{\partial q^{(2)}} \right )=(\omega_{1}^{2}+\omega_{2}^{2})\dot{x}_{1}-\dddot{x}_{1},\\
P_{2}&=\ddot{x}_{1}.
\end{split}
\end{equation}

The Hamiltonian can be obtained from Eq.\eqref{E13}:
\begin{equation}
    \label{E32}
    H=P_{1}Q_{2}+\frac{P_{2}}{2}^{2}-\frac{1}{2}\left[(\omega_{1}^{2}+\omega_{2}^{2})Q^{2}_{2}+\omega_{1}^{2}\omega_{2}^{2}Q_{1}^{2}\right].
\end{equation}
It is important to emphasize that the Hamiltonian of Eq.~(31) is not bounded from below, since it is linear in $P_{1}$. This feature reflects the well-known Ostrogradsky instability, which has been extensively discussed in the literature in the context of higher-order theories. In particular, the Pais--Uhlenbeck oscillator has served as a toy model to explore such instabilities and the appearance of so-called ghost degrees of freedom \cite{woodard2015theorem,pavvsivc2016pais,svanberg2022theories}.

Using the Hamilton-Ostrogradsky equations from Eq.\eqref{E14}, we have:
\begin{equation}
\begin{split}
\label{E33}
\dot{Q}_{1}&=\frac{\partial H}{\partial P_{1}}=Q_{2},\\
\dot{Q}_{2}&=\frac{\partial H}{\partial P_{2}}=P_{2},\\
\dot{P}_{1}&=-\frac{\partial H}{\partial Q_{1}}=\omega_{1}^{2}\omega_{2}^{2}Q_{1},\\
\dot{P}_{2}&=-\frac{\partial H}{\partial Q_{2}}=-P_{1}+(\omega_{1}^{2}+\omega_{2}^{2})Q_{2}.
\end{split}\,
\end{equation}

The solutions to these equations are given by:
\begin{equation}
    \label{E34}
    \begin{aligned}
    Q_{1}(t)=A_{1}\cos(\omega_{1}t)+B_{1}\sin(\omega_{1}t)\\
+A_{2}\cos(\omega_{2}t)+B_{2}\sin(\omega_{2}t),
\end{aligned}
\end{equation}
\begin{equation}
    \label{E35}
    \begin{aligned}
   Q_{2}(t)=\omega_{1}\left[-A_{1}\sin(\omega_{1}t)+B_{1}\cos(\omega_{1}t)\right]\\
   +\omega_{2}\left[-A_{2}\sin(\omega_{2}t)+B_{2}\cos(\omega_{2}t)\right],
   \end{aligned}
\end{equation}
\begin{equation}
    \label{E36}
    \begin{aligned}
   P_{1}(t)=\left[-A_{1}\sin(\omega_{1}t)+B_{1}\cos(\omega_{1}t)\right]\omega_{1}(2\omega_{1}^{2}+\omega_{2}^{2})\\
   +\left[-A_{2}\sin(\omega_{2}t)+B_{2}\cos(\omega_{2}t)\right]\omega_{2}(\omega_{1}^{2}+2\omega_{2}^{2}),
\end{aligned}
\end{equation}
\begin{equation}
    \label{E37}
    \begin{aligned}
    P_{2}(t)=\left[-A_{1}\cos(\omega_{1}t)-B_{1}\sin(\omega_{1}t)\right]\omega_{1}^{2}\\
    +\left[-A_{2}\cos(\omega_{2}t)-B_{2}\sin(\omega_{2}t)\right]\omega_{2}^{2}.
\end{aligned}
\end{equation}

These solutions are easily obtained knowing that $Q_{1}(t)=x_{1}(t)$, simply by using the relations that appear in Eq.\eqref{E33}. Similarly, the same can be done for $x_{2}(t)$, although we do not present it here since the results are similar and there is no need to repeat the mathematical procedures.

Note that these solutions satisfy the constraint given by Eq.~\eqref{E12}. A more direct calculation can be carried out by writing
\begin{equation*}
    P_1+\dot{P_2}-\frac{\partial L}{\partial \dot{x}}
    =(\omega_{1}^{2}+\omega_{2}^{2})Q_{2}-(\omega_{1}^{2}+\omega_{2}^{2})\dot{x}=0
\end{equation*}
where we have used the last equation of \eqref{E33} together with the fact that $Q_2=\dot{x}$. This means that the constraints arising from the Hamilton-Ostrogradsky formulation are already satisfied in Eqs. \eqref{E34} to \eqref{E37}.

Finally, it can be said that the Pais-Uhlenbeck oscillator model is not so simple to visualize in a first study, as we are not used to dealing with energies that depend on accelerations as seen in $L_{\mathrm{PU}}$. However, the coupled oscillator model is easier to interpret and visualize. Due to the equivalence that coupled oscillators have with the Pais-Uhlenbeck oscillator, it becomes a good model to be given as an example in an initial approach to higher-order equations.

This work serves as a brief introduction to the Ostrogradsky formulation, emphasizing key aspects such as the Lagrange and Hamilton equations for higher-order systems. The interested reader can consult references \cite{Rashid1996, grosse1993effective, woodard2015theorem, svanberg2022theories} for a deeper understanding of the Ostrogradsky formulation for higher-order equations. For a more in-depth reading on the Pais-Uhlenbeck oscillator, refer to \cite{Mendes2017,pavvsivc2016pais}.

For the reader interested in practicing Ostrogradski's method, we have provided a suggested exercise in Appendix C, the so-called \textit{Snap Oscillator}, a toy model whose Lagrangian depends on the acceleration squared \cite{woodard2015theorem}. This system leads to a fourth-order equation of motion in a straightforward way, making it suitable as an exercise for students to practice the use of higher-order Euler–Lagrange equations and the Ostrogradsky formalism. For this reason, we propose it as a pedagogical complement to the Pais–Uhlenbeck oscillator, providing a simpler entry point to the study of higher-order dynamics.

\section{Conclusion}
In this work, we provide a brief introduction to higher-order equations of motion. We then introduced the approach used by Ostrogradski to transition from the Lagrangian formalism with higher-order derivatives to the Hamiltonian formalism, which is known as the Hamilton-Ostrogradski formalism. Finally, we presented the model of two coupled oscillators and the Pais-Uhlenbeck model, whose Lagrangians are equivalent. The Lagrangian of the Pais-Uhlenbeck oscillator depends on second-order derivatives, which is not common in physics. On the other hand, the Lagrangian of two coupled oscillators depends on derivatives up to the first order and is a model that is easier to interpret.

It is an interesting issue that Nature, in general, is as far as we know described by laws that only involve derivatives up to second order. This issue has motivated several scholars to seek physical situations described by higher-order derivatives. It is believed that if such theoretical models can be experimentally proven, this would expand our understanding of nature.

We believe that the approach presented here can serve as an introductory basis to spark the curiosity of students interested in delving deeper into the Lagrangian and Hamiltonian formulations in the higher-order regime. Furthermore, this study can be extended in future works to address topics such as generalizations of Poisson brackets, conservation laws, Noether’s theorem, Liouville’s equation, and symplectic group symmetries in higher-order systems.

As a result, we hope that this work may help to stimulate interest in higher-order systems and encourage further studies that consolidate their role both in teaching and in research in theoretical physics.

\section*{Acknowledgements}

CAMM is grateful to FAPEMIG-Brazil (Grants APQ-00544- 23 and APQ-05218-23) for partial financial support and to DFA of UniCT and INFN-LNS-CT for the hospitality. This study was financed in part by the Coordenação de Aperfeiçoamento de Pessoal de Nível Superior – Brazil (CAPES) – Finance Code 001, Process No. 88887.820319/2023-00. Both authors are very grateful to the reviewers and editors, whose comments and suggestions helped to greatly improve this work.

\section*{Appendix}
\subsection*{Appendix A: Higher-Order Lagrange Equations}
For simplicity, let us assume that the system has only one degree of freedom and that the Lagrangian does not explicitly depend on time. Consider the Lagrangian $L=L(q,q^{(1)},q^{(2)})$, where we use the notation $q^{(n)}=d^{n}q/dt^{n}$ to indicate the $n$-th temporal derivative of the generalized coordinate. Thus, $q^{(1)}$ and $q^{(2)}$ are the first and second-order derivatives, respectively.

From the principle of least action, we can obtain the Lagrange equation for this type of Lagrangian:
\begin{equation}
\label{A2}
\delta S=\delta \int_{t_{1}}^{t_{2}}L(q,q^{(1)},q^{(2)})\mathrm{d}t=0.
\end{equation}

By applying the variation to the right-hand side of Eq.\eqref{A2}, we get:
\begin{equation}
\label{A3}
\int_{t_{1}}^{t_{2}}\left ( \frac{\partial L}{\partial q}\delta q+\frac{\partial L}{\partial \dot{q}}\delta \dot{q}+\frac{\partial L}{\partial \ddot{q}}\delta \ddot{q} \right )\mathrm{d}t=0.
\end{equation}

By imposing the condition that $\delta q(t_{1})=\delta q(t_{2})=\delta \dot{q}(t_{1})=\delta \dot{q}(t_{2})=0$, we can perform integration by parts to eliminate $\delta \dot{q}$ and $\delta \ddot{q}$:
\begin{equation}
    \label{A4}
    \begin{aligned} 
\int_{t_{1}}^{t_{2}} \frac{\partial L}{\partial \dot{q}}\, \delta \dot{q} \,\mathrm{d}t 
&= \left. \frac{\partial L}{\partial \dot{q}}\, \delta q \right|_{t_{1}}^{t_{2}} 
- \int_{t_{1}}^{t_{2}} \frac{\mathrm{d}}{\mathrm{d}t} \left( \frac{\partial L}{\partial q} \right) \delta q \,\mathrm{d}t =\\
&= -\int_{t_{1}}^{t_{2}} \frac{\mathrm{d}}{\mathrm{d}t} \left( \frac{\partial L}{\partial q} \right) \delta q \,\mathrm{d}t.
\end{aligned}
\end{equation}

\begin{equation}
    \label{A5}
    \begin{aligned}
\int_{t_{1}}^{t_{2}} \frac{\partial L}{\partial \ddot{q}}\, \delta \ddot{q} \,\mathrm{d}t 
&= \left. \frac{\partial L}{\partial \ddot{q}}\, \delta \dot{q} \right|_{t_{1}}^{t_{2}} 
- \int_{t_{1}}^{t_{2}} \frac{\mathrm{d}}{\mathrm{d}t} \left( \frac{\partial L}{\partial \ddot{q}} \right) \delta \dot{q} \,\mathrm{d}t =\\
&= -\int_{t_{1}}^{t_{2}} \frac{\mathrm{d}}{\mathrm{d}t} \left( \frac{\partial L}{\partial \ddot{q}} \right) \delta \dot{q} \,\mathrm{d}t =\\
&= -\left. \frac{\mathrm{d}}{\mathrm{d}t} \left( \frac{\partial L}{\partial \ddot{q}} \right) \delta q \right|_{t_{1}}^{t_{2}} 
+ \int_{t_{1}}^{t_{2}} \frac{\mathrm{d}^2}{\mathrm{d}t^2} \left( \frac{\partial L}{\partial \ddot{q}} \right) \delta q \,\mathrm{d}t =\\
&= \int_{t_{1}}^{t_{2}} \frac{\mathrm{d}^2}{\mathrm{d}t^2} \left( \frac{\partial L}{\partial \ddot{q}} \right) \delta q \,\mathrm{d}t.
    \end{aligned}
\end{equation}

Substituting the results from Eq.\eqref{A4} and \eqref{A5} into Eq.\eqref{A3}, we get:
\begin{equation}
    \label{A6}
    \int_{t_{1}}^{t_{2}}\left ( \frac{\partial L}{\partial q}-\frac{\mathrm{d} }{\mathrm{d} t}\frac{\partial L}{\partial \dot{q}}+\frac{\mathrm{d}^{2} }{\mathrm{d} t^{2}}\frac{\partial L}{\partial \ddot{q}} \right )\delta q\mathrm{d}t=0.
\end{equation}

Since $\delta q$ is an arbitrary variation, we extract the Lagrange equation:
\begin{equation}
    \label{A7}
    \frac{\partial L}{\partial q}-\frac{\mathrm{d} }{\mathrm{d} t}\left(\frac{\partial L}{\partial \dot{q}}\right)+\frac{\mathrm{d}^{2} }{\mathrm{d} t^{2}}\left(\frac{\partial L}{\partial \ddot{q}}\right)=0.
\end{equation}

To generalize the Lagrange equation to higher orders, consider a Lagrangian that depends on derivatives up to order $n$:
\begin{equation}
    \label{A8}
    L=L(q,q^{(1)},q^{(2)},...,q^{(n)}).
\end{equation} 

To obtain the higher-order Lagrange equations, it is necessary to impose that \footnote{When the Lagrangian has the form \(L(q, \dot{q}, t)\), one must to impose the conditions \(\delta q(t_1) = \delta q(t_2) = 0\) in order to obtain the Lagrange equations. If the Lagrangian depends on higher-order derivatives, \(L(q, \dot{q}, \ddot{q}, \ldots, q^{(n)}, t)\), analogous conditions must also be imposed for the variations of the derivatives up to order \(n-1\), namely \(\delta q^{(k)}(t_1) = \delta q^{(k)}(t_2) = 0\) for \(k = 0, 1, \ldots, n-1\).
}:
\begin{equation}
    \label{A9}
\delta q^{(k)}(t_{1})=\delta q^{(k)}(t_{2})=0 \quad (k=1,...,n-1).
\end{equation}

This will allow integration by parts such that only $\delta q$ coefficients appear. Following similar steps to the $n=2$ case, the higher-order Lagrange equation is:
\begin{equation}
    \label{A10}
    \sum_{k=0}^{n}(-1)^{k}\frac{\mathrm{d}^{k} }{\mathrm{d} t^{k}}\left(\frac{\partial L}{\partial q^{(k)}}\right)=0.
\end{equation}

For a system with $N$ degrees of freedom, there will be a set of $N$ generalized coordinates and $N$ higher-order Lagrange equations:
\begin{equation}
    \label{A10}
    \sum_{k=0}^{n}(-1)^{k}\frac{\mathrm{d}^{k} }{\mathrm{d} t^{k}}\left(\frac{\partial L}{\partial q_{i}^{(k)}}\right)=0 \quad (i=1,...,N).
\end{equation}

Here it is assumed that the $N$ generalized coordinates are all independent, meaning that the $\delta q_{i}$ are mutually independent and arbitrary.

\subsection*{Appendix B: Higher-Order Hamilton Equations}
In Appendix A we showed how to derive the Lagrange equation for a Lagrangian depending on up to second-order derivatives, and then, with some considerations, presented the general form of the Lagrange equation for derivatives up to order $n$.

In this appendix, we will show how to derive Hamilton’s equations for a Lagrangian depending on up to second-order derivatives, and then present the generalization of Hamilton’s equations to the $n$-th order.

It is known that to switch from the Lagrangian formalism to the Hamiltonian formalism one uses the Legendre transformation. In the common case where the Lagrangian is $L(q_{i},\dot{q}_{i},t)$, the Legendre transformation takes the form:
\begin{equation}
    \label{B1}
    H(q_{i},p_{i},t)=\sum_{i=1}^{N}(p_{i}\dot{q}_{i})-L(q_{i},\dot{q}_{i},t),
\end{equation}
where $p_{i}$ is the conjugate momentum, defined as:
\begin{equation}
    \label{B2}
    p_{i}=\frac{\partial L}{\partial \dot{q}_{i}} \quad (i=1,\dots,N).
\end{equation}

Let the Hessian matrix be formed by elements $W_{ij}$, given by:
\begin{equation}
\label{B3}
    W_{ij}=\frac{\partial^2 L}{\partial \dot{q}_{i}\,\partial\dot{q}_{j}} \quad (i,j=1,\dots,N).
\end{equation}

If the Hessian matrix is regular, i.e., has nonzero determinant, then one can solve the $N$ equations of Eq.~\eqref{B2} to express the $\dot{q}_{i}$ as functions of $q_{i}$, $p_{i}$, and $t$. Substituting these expressions into Eq.~\eqref{B1} yields the Hamiltonian.

Hamilton’s equations form a set of $2N$ first-order differential equations:
\begin{equation}
\label{B5}
\dot{q}_{i}=\frac{\partial H}{\partial p_{i}} ,\quad \dot{p}_{i}=-\frac{\partial H}{\partial q_{i}} \quad (i=1,\dots,N).
\end{equation}

Like the Lagrange equations, Hamilton’s equations yield the equations of motion, but whereas the Lagrange equations are second order, Hamilton’s equations are first order. The Lagrangian formalism deals with $N$ equations in $N$ generalized coordinates $(q_{1},\dots,q_{N})$, while the Hamiltonian formalism involves $2N$ equations in the $2N$ variables $(q_{1},\dots,q_{N},p_{1},\dots,p_{N})$.

When considering higher-order derivatives, the Legendre transformation as given in Eq.~\eqref{B1} is no longer sufficient. One must generalize the Legendre transformation to handle new coordinates and momenta.

For simplicity, let us again consider a system with a single degree of freedom. If the Lagrangian is of the type $L(q,\dot{q},\ddot{q})$, the Lagrange equation reads:
\begin{equation}
    \label{B5}
    \frac{\partial L}{\partial q}-\frac{\mathrm{d}}{\mathrm{d}t}\left(\frac{\partial L}{\partial \dot{q}}\right)+\frac{\mathrm{d}^{2}}{\mathrm{d}t^{2}}\left(\frac{\partial L}{\partial \ddot{q}}\right)=0.
\end{equation}

New canonical variables are then defined as:
\begin{equation}
\label{B6}
\begin{aligned}
    Q_{1} &= q, & \quad Q_{2} &= \dot{q}, \\[6pt]
    P_{1} &= \frac{\partial L}{\partial \dot{q}} 
            - \frac{\mathrm{d}}{\mathrm{d}t}\!\left(\frac{\partial L}{\partial \ddot{q}}\right), 
    & \quad P_{2} &= \frac{\partial L}{\partial \ddot{q}}.
\end{aligned}
\end{equation}

where $Q_{1}$ and $Q_{2}$ are the new generalized coordinates and $P_{1}$ and $P_{2}$ are the corresponding momenta. Note that these definitions introduce constraints, for example:
\begin{equation}
\label{B7}
\frac{\partial L}{\partial \dot{q}}=P_{1}+\dot{P}_{2}.
\end{equation}

Hence, the coordinates and momenta are not all independent. Because $P_{2}$ depends on $q$, $\dot{q}$, and $\ddot{q}$, to invert this relation and express $\ddot{q}$ as a function of $q$, $\dot{q}$, and $P_{2}$, the Hessian must be regular. In this higher-order case the Hessian matrix is formed by derivatives with respect to the generalized accelerations:
\begin{equation}
\label{B8}
W_{ij}=\frac{\partial^2 L}{\partial \ddot{q}_{i}\,\partial\ddot{q}_{j}} \quad (i,j=1,\dots,N).
\end{equation}

In the one-dimensional case, this reduces to a single element:
\begin{equation}
\label{B9}
    W=\frac{\partial^2 L}{\partial \ddot{q}^{2}}.
\end{equation}

If $\det(W)\neq 0$, one can write:
\begin{equation}
\label{B10}
    \ddot{q}=\ddot{q}(q,\dot{q},P_{2})=\ddot{q}(Q_{1},Q_{2},P_{2}).
\end{equation}

To derive the Legendre transformation, start from the differential of the Lagrangian:
\begin{equation}
\label{B11}
    \mathrm{d}L=\frac{\partial L}{\partial q}\,\mathrm{d}q+\frac{\partial L}{\partial \dot{q}}\,\mathrm{d}\dot{q}+\frac{\partial L}{\partial \ddot{q}}\,\mathrm{d}\ddot{q}.
\end{equation}

Using the definitions in Eq.~\eqref{B6}, this becomes:
\begin{equation}
    \label{B12}
    \mathrm{d}L=\dot{P}_{1}\,\mathrm{d}Q_{1}+(P_{1}+\dot{P}_{2})\,\mathrm{d}Q_{2}+P_{2}\,\mathrm{d}\ddot{q}.
\end{equation}

To obtain only differentials of $Q_{i}$ and $P_{i}$, apply the identity $\ddot{q}\,dP = d(P\,\ddot{q}) - P\,d\ddot{q}$, yielding:
\begin{equation}
\label{B13}
\mathrm{d}L=\mathrm{d}\bigl(P_{1}Q_{2}+\ddot{q}\,P_{2}\bigr)+\dot{P}_{1}\,\mathrm{d}Q_{1}-Q_{2}\,\mathrm{d}P_{1}+\dot{P}_{2}\,\mathrm{d}Q_{2}-\ddot{q}\,\mathrm{d}P_{2}.
\end{equation}

Equation \eqref{B13} can now be used to define the differential of the Hamiltonian function for a higher order theory:
\begin{equation}
\label{B14}
    \begin{aligned}
\mathrm{d}H&=\mathrm{d}\bigl(P_{1}Q_{2}+\ddot{q}\,P_{2}-L\bigr)=\\
    &=-\dot{P}_{1}\,\mathrm{d}Q_{1}+Q_{2}\,\mathrm{d}P_{1}-\dot{P}_{2}\,\mathrm{d}Q_{2}+\ddot{q}\,\mathrm{d}P_{2}.
    \end{aligned}
\end{equation}

This last equation provides the precise expression for the differential of the Hamiltonian in terms of the canonical variables. The relevance of this formulation lies in ensuring that the formalism preserves the canonical structure of Hamiltonian mechanics, allowing the time evolution to be described through the generalized Hamilton equations. With this result in hand, the Legendre transformation can be written as:
\begin{equation}
\label{B15}
H(Q_{1},Q_{2},P_{1},P_{2})=P_{1}Q_{2}+P_{2}\ddot{q}-L(Q_{1},Q_{2},\ddot{q}).
\end{equation}
where $\ddot{q}=\ddot{q}(Q_{1},Q_{2},P_{2})$. Note also that
\begin{equation}
\label{B16}
    Q_{2}=\frac{\mathrm{d}q}{\mathrm{d}t}=\dot{Q}_{1},\quad \ddot{q}=\frac{\mathrm{d}\dot{q}}{\mathrm{d}t}=\dot{Q}_{2}
\end{equation}

Hence, for a Lagrangian depending on $\ddot{q}$, the Legendre transformation reads:
\begin{equation}
\label{B17}
H=P_{1}\dot{Q}_{1}+P_{2}\dot{Q}_{2}-L.
\end{equation}

From Eq.~\eqref{B14} we then read off Hamilton’s equations:
\begin{equation}
    \label{B18}
    \dot{Q}_{1}=\frac{\partial H}{\partial P_{1}},\quad \dot{Q}_{2}=\frac{\partial H}{\partial P_{2}},\quad \dot{P}_{1}=-\frac{\partial H}{\partial Q_{1}},\quad \dot{P}_{2}=-\frac{\partial H}{\partial Q_{2}}.
\end{equation}

Note that Eq.~\eqref{B5} is a single fourth-order equation in one variable, whereas Eq.~\eqref{B18} is a system of four first-order equations in four variables.

For the general case of a Lagrangian depending on derivatives up to order $n$:
\begin{equation}
     L=L(q,q^{(1)},q^{(2)},\dots,q^{(n)}),
\end{equation}
with $q^{(1)}=\dot{q}$, $q^{(2)}=\ddot{q}$, etc., one defines:
\begin{equation}
     Q_{k}=q^{(k-1)} \quad(k=1,\dots,n)
\end{equation}
and
\begin{equation}
    P_{k} = \sum_{m=k}^{n} (-1)^{m-k} \frac{\mathrm{d}^{m-k}}{\mathrm{d}t^{m-k}} 
    \left( \frac{\partial L}{\partial q^{(m)}} \right)
    \qquad (k = 1, \dots, n).
\end{equation}

The general Legendre transformation is:
\begin{equation}
 H=\sum_{k=1}^{n}P_{k}\dot{Q}_{k}-L,
\end{equation}
and the higher-order Hamilton equations take the form:
\begin{equation}
    \dot{Q}_{k}=\frac{\partial H}{\partial P_{k}},\quad \dot{P}_{k}=-\frac{\partial H}{\partial Q_{k}}\quad(k=1,\dots,n).
\end{equation}

\subsection*{Appendix C: Proposed Exercise}

This appendix presents a didactic exercise designed for the student to apply and consolidate some of the concepts discussed in this work. We introduce here a simple model that we call the \textit{Snap Oscillator} \cite{woodard2015theorem}, since its equation of motion involves time derivatives up to fourth order. Its Lagrangian is defined as
\begin{equation}
    L(x,\dot{x},\ddot{x}) = -\frac{\epsilon m}{2\omega^{2}}\ddot{x}^{2} + \frac{m}{2}\dot{x}^{2} - \frac{m\omega^{2}}{2}x^{2},
\end{equation}
where $\epsilon > 0$ is a constant parameter.

\medskip

\begin{enumerate}
    \item Using the higher-order Euler–Lagrange equation, derive the equation of motion for the Snap Oscillator.
    \item Solve the resulting equation of motion. Discuss the limit $\epsilon \to 0$, and verify explicitly whether the solution reduces to that of the simple harmonic oscillator. Note that the most appropriate way to do this limit is using the \textit{Singular Perturbation Theory} \cite{Chen1995RenormalizationGA, OMalley1991SingularPM}.
    \item Apply Ostrogradsky’s formalism to this problem. In particular, determine the canonical variables:
    \[
    \begin{aligned}
    Q_1 &= x, & \quad Q_2 &= \dot{x}, \\[6pt]
    P_1 &= \frac{\partial L}{\partial \dot{x}} - \frac{d}{dt}\!\left(\frac{\partial L}{\partial \ddot{x}}\right), 
    & \quad P_2 &= \frac{\partial L}{\partial \ddot{x}}.
    \end{aligned}
    \]
    Compute $P_1$ and $P_2$ explicitly.
    \item Construct the Hamiltonian $H(Q_1,Q_2,P_1,P_2)$ using the generalized Legendre transformation.
\end{enumerate}

\bibliography{Referencias}

\end{document}